\documentclass[11pt,a4paper]{article}

\usepackage{subfigure}
\usepackage{amssymb}
\usepackage{xcolor}
\usepackage{graphicx}
\usepackage{float}
\usepackage{amssymb}
\usepackage{amsmath}
\usepackage{amsfonts}

\newcommand{\cwhile}{\textbf{while }}
\newcommand{\cendwhile}{\textbf{end while }}
\newcommand{\cdo}{\textbf{do }}
\newcommand{\cif}{\textbf{if }}
\newcommand{\cendif}{\textbf{end if }}
\newcommand{\celse}{\textbf{else }}

\begin{document}




\title{Identifying the number of clusters in discrete mixture models}

\author{Cl\'{a}udia Silvestre,$^{\rm a}$ $^{\ast}$
\vspace{6pt} Margarida G. M. S. Cardoso,$^{\rm b}$ \vspace{6pt} M\'{a}rio A. T. Figueiredo$^{\rm c}$ \\ \vspace{6pt}
$^{a}${\small{\em{Escola Superior de Comunica\c{c}\~{a}o Social, Instituto Polit\'{e}cnico de Lisboa, Portugal}}};\\
$^{b}${\small{\em{Department of Quantitative Methods, ISCTE - Lisbon University Institute, Portugal}}};\\
$^{c}${\small{\em{Instituto de Telecomunica\c{c}\~{o}es, Instituto Superior T\'{e}cnico, Universidade de Lisboa, Portugal,}}}
}

 \maketitle

\begin{abstract}
Research on cluster analysis for categorical data continues to develop,
with new clustering algorithms being proposed. However, in this context, the determination of the number of clusters is rarely addressed. In this paper, we propose a new approach in which clustering of  categorical data and the
estimation of the number of clusters is carried out  simultaneously.
Assuming that the data originate from a finite mixture of multinomial
distributions, we develop a method to select the number of mixture components
based on a {\it minimum message length} (MML) criterion
and implement a new {\it expectation-maximization} (EM) algorithm
to estimate all the model parameters.
The proposed EM-MML approach, rather than selecting one among a set
of pre-estimated candidate models (which requires running EM several
times), seamlessly integrates estimation and model selection in a
single algorithm.
The performance of the proposed
approach is compared with other well-known criteria (such as the {\it Bayesian information criterion}--BIC), resorting to synthetic data and to two real  applications from the European Social Survey. The  EM-MML computation time is  a clear advantage of the proposed method.  Also, the real data solutions are much more parsimonious than the solutions provided by competing methods, which reduces the risk of model order overestimation
and increases interpretability.

\medskip 

\textbf{keywords}: finite mixture model; EM algorithm; model selection, minimum message length; categorical data
%
%
\end{abstract}

\section{Introduction}

Clustering is a technique commonly used in several research and application areas, such as social sciences,
medicine, biology, engineering, computer science, image analysis, bioinformatics, and marketing.
The goal of clustering is to discover or uncover groups in data. To this end,
there are essentially two different approaches: distance-based,
where a distance or a similarity measure between objects is defined and similar objects are assigned to the same group; model-based, where the data are assumed to be generated by a finite mixture model, and objects are assigned to groups based on the corresponding estimates of the posterior probabilities \cite{CZLMW2012}.

Most of the clustering techniques are focused on numerical data and can not be applied directly to categorical data.
In fact, clustering techniques for categorical data are more challenging  \cite{XWMM2012}, and there are fewer techniques available \cite{GiordanDiana2011}.
When using distance-based clustering approaches, one needs to resort to specific similarity measures to deal with categorical features -- e.g \cite{ChenLiu2009}. In this context, the determination of the number of clusters is commonly based on clustering quality measures and the corresponding graphical displays, such as dendrograms (when using hierarchical clustering techniques) or entropy related graphics  \cite{DesaiSinghPudi2011}. In model-based approaches for numerical data, the usual choice is a mixture of Gaussians (e.g. \cite{CTL2006}, \cite{COL2005}, \cite{LFJ2004}); when referring to  categorical data, a mixture of multinomials -- discrete mixture model -- is usually considered (e.g. \cite{GiordanDiana2011}, \cite{Bouguila2008}, \cite{LiZhang2008}, \cite{DeyaLim2013}). In order to determine the number of groups in discrete mixture models,
information criteria are commonly used: e.g., the {\it Bayesian information criterion} (BIC) \cite{Schwarz1978}  or  the {\it Akaike information criterion} (AIC) \cite{Akaike1973}. These criteria look for a balance between the model's fit to the data (which ocrresponds to maximizing the likelihood function) and parsimony (using penalties associated with measures of model complexity), thus trying to avoid over-fitting. The use of information criteria follows the estimation of candidate finite mixture models for which a predetermined  number of groups is indicated, generally resorting to an EM ({\it expectation-maximization}) algorithm, \cite{DLR1997}.

In this work, we focus on determining the number of groups while clustering categorical data,
using an EM embedded approach  to estimate the number of groups.
The novelty of this approach is that it does not rely on selecting among a set of pre-estimated candidate models, but rather integrates estimation and model selection in a single algorithm. We capitalize on the approach developed by Figueiredo and Jain \cite{FJ2002} for clustering continuous data and extend it for dealing with categorical data.
The proposed method is based on a {\it minimum message length} (MML) criterion to select the number of clusters and on an EM algorithm to estimate the model parameters; our implementation follows a previous version described in \cite{SCF2008}.

The paper is organized as follows: Section 2 reviews the finite mixture model-based approach for clustering
and addresses the case of categorical data; Section 3 provides an introduction to the topic of model selection for discrete finite mixtures; in Section 4 we describe the proposed EM-MML based algorithm; in Section 5, the experimental results, based on synthetic and real data, illustrate the performance of the EM-MML approach. Concluding remarks are summarized in Section 6.

\section{Clustering with finite mixture models}

Finite mixture models  offer a model-based  approach to clustering, exhibiting some competitive advantages when compared to alternative methods:
besides producing the allocation of observations to clusters,  they yield estimates of within-clusters joint probability functions for the base variables; moreover,
they provide means to allocate new observations to groups; finally,
when used with information criteria, mixture models provide a statistical framework to
determine the number of clusters.

The literature on finite mixture models and their application is vast, including some books  covering theory, geometry, and  applications \cite{EverittHand1981}, \cite{TitteringtonSmithMakov1985}, \cite{McLachlanPeel2000}, \cite{TSM1985}, \cite{MelnykovMaitra2010}, \cite{Lindsay1995}.
For example, in market segmentation,  cluster analysis via
finite mixture models has replaced more traditional cluster analysis, such as K-Means algorithm, as the state of the art \cite{WedelKamakura2000}.

Finite mixture models have played an important role in the history of modern statistics. One of the first applications of mixture models is due to
Newcomb \cite{Newcomb1886}, who used a mixture of Gaussians to analyse a collection
of observations of transits of Mercury. Pearson  \cite{Pearson1894} fitted a mixture of two Gaussians with unequal variances in an analysis of
different species of crabs. Among many other examples of applying mixture models,
MacDonald and Pitcher \cite{MacdonaldPitcher1979} analysed single species of fish in a lake,  using a mixture of Gaussians
where each component consists of the fish of a single yearly spawning of that species.
Another example is given by Do and McLachlan \cite{DoMcLachlan1984},
where mixture of Gaussians was used to study the populations of rats that were being eaten by a particular species of owl, with the distinct rat species corresponding to the components of mixtures. Al-Hussaini and Abdel-Hamid \cite{HussainiHamid2006} studied the behavior of failure time of a device; they fitted a mixture of components, each of which represents a different cause of failure. In their research, a special attention was paid to mixtures of two Weibull components, but two exponential components, two Rayleigh components,  and
mixture of Rayleigh and Weibull components, were also analysed.
In these examples, the segmentation process uncovers physically meaningful components.

When applying  finite mixture models to social sciences, the
analyst may be confronted with the need to uncover sub-populations
based on qualitative indicators. In this context, the use of mixture models of categorical data is particularly pertinent.
For example, for clustering categorical data from multiple choice questions,
a mixture of multinomial distributions is used in
order to market products \cite{MaitraMelnykov2010b}.

\subsection{Definitions and concepts}
Let $ \textbf{Y} = \{ \underline{y}_i,  \;    i=1, \dots, n\}$
be a set of $n$ independent and identically distributed (i.i.d.)
sample observations of a random vector, $ \underline{Y}=[Y_1,\dots, Y_L ]'$.
If $\underline{Y}$ follows a mixture of $K$ components densities,
$f(\underline{y}|\underline{\theta}_k)$ $(k=1, \dots, K)$, with
probabilities $\{ \alpha_1, \dots,\alpha_K \}$, the
probability (density) function of $\underline{Y}$ is
\begin{displaymath}
f(\underline{y}|\Theta) = \sum_{k=1}^{K} \alpha_k f(\underline{y}|\underline{\theta}_k),
\end{displaymath}
where $\Theta=\{\underline{\theta}_1,\dots,\underline{\theta}_K,\alpha_1,\dots,\alpha_K \}$ is the set of all the parameters of the model and $\underline{\theta}_k$ are the distributional parameters defining the $k$-{th} component.
The proportions, also called {\it mixing probabilities}, are subject to the usual constraints:
$\sum_{k=1}^K \alpha_k = 1 $ and $\alpha_k \geq 0$, $k=1,\dots,K$.

The log-likelihood of the observed set of sample observations is
\begin{displaymath}
\log f(\textbf{Y}|\Theta) =\log \prod_{i=1}^n{f(\underline{y}_i|\Theta)}=\sum_{i=1}^{n}\log \sum_{k=1}^{K} \alpha_k f(\underline{y}_i|\underline{\theta}_k).
\end{displaymath}
In clustering, the identity of the component that generated each sample observation is unknown. The observed data $\textbf{Y}$ is therefore regarded as incomplete, where the missing data is a set of indicator variables $\textbf{Z} = \{\underline{z}_1,...,\underline{z}_n\}$, each taking the form $\underline{z}_i = [z_{i1},...,z_{iK}]'$, where $z_{ik}$ is a binary indicator: $z_{ik}$ takes the value 1 if the observation
$\underline{y}_i$ was generated by the k-th component, and $0$ otherwise.
It is usually assumed that the $\{\underline{z}_i, \; i=1, \dots, n\}$ are i.i.d.,
following a multinomial distribution of $K$ categories, with probabilities $\{\alpha_1,\dots,\alpha_K \}$. The log-likelihood of complete data
$\{\textbf{Y},\textbf{Z} \}$ is given by
\begin{equation*}
 \log f( \textbf{Y},\textbf{Z}|\Theta) = \sum_{i=1}^n \sum_{k=1}^K z_{ik} \log
 \left[ \alpha_k
 f(\underline{y}_i|\underline{\theta}_k) \right].
 \label{eq:complete}
\end{equation*}

\subsection{Discrete Finite Mixture Models}
Consider that each variable in \underline{Y}, $Y_l$ ($l=1,\dots,L$) can take one of $C_l$ categories. Conditionally on having been generated by the k-th component of the  mixture, each $Y_{l}$ is thus modeled by a multinomial distribution with $n_l$ trials, $C_l$ categories, and non-negative parameters $\underline{\theta}_{kl} = \{\theta_{klc},\; c=1,\dots,C_l\}$, with $\sum_{c=1}^{C_l} \theta_{klc} = 1$.
For a sample $y_{il} (i=1, \dots,n)$ of $Y_l$, we denote as $y_{ilc}$ the number of outcomes in category $c$, which is a sufficient statistic; naturally, $\sum_{c=1}^{C_l} y_{ilc} = n_l$.
Thus, with $\underline{\theta}_k = \{\underline{\theta}_{k1},\dots,\underline{\theta}_{kL}\}$
and $\Theta = \{\underline{\theta}_1,\dots,\underline{\theta}_K,\alpha_1,\dots,\alpha_k\}$,
the log-likelihood function, for a set of observations corresponding to a discrete finite mixture model (mixture of multinomials) is
\begin{equation}
 \label{mixMultinomial}
\log p(\textbf{Y}|\Theta) = \sum_{i=1}^n \log
\sum_{k=1}^K \alpha_k \prod_{l=1}^L \left[ n_l! \prod_{c=1}^{C_l} \frac{(\theta_{klc})^{y_{ilc}}}{y_{ilc}!} \right].
\end{equation}

In order to estimate the parameters of this mixture, it is necessary to ensure that they are identifiable. In particular, in the case of a mixture of multinomials, specific identifiability condition must be fulfilled: a mixture of multinomial distributions is identifiable if $T\geq2K-1$, where $T$ is the number of trials of each multinomial distribution (see details in \cite{ElmoreWang2003} and \cite{Portela2008}).

As in the Gaussian case, the log-likelihood in \eqref{mixMultinomial} can be seen as corresponding to a missing-data problem, where the missing data has exactly the same meaning and structure as above. The log-likelihood of the complete data
$\{\textbf{Y},\textbf{Z} \}$ is thus given by

\begin{equation}
 \label{mixMultinomialcomplete}
\log p(\textbf{Y},\textbf{Z}|\Theta) = \sum_{i=1}^n
\sum_{k=1}^K z_{ik} \log \left(\alpha_k \prod_{l=1}^L \left[ n_l! \prod_{c=1}^{C_l} \frac{(\theta_{klc})^{y_{ilc}}}{y_{ilc}!} \right]\right).
\end{equation}


\subsection{Estimation of finite mixture models via the EM algorithm}
To obtain a {\it maximum-likelhood} (ML) or {\it maximum a posteriori} (MAP) estimate of the parameters of a multinomial mixture, the well-known EM algorithm is usually the tool of choice (\cite{DLR1997} and \cite{VermuntMagidson2002}).
EM is an iterative algorithm, which alternates between two steps, the {\it expectation step} (E-step) and the {\it maximization step} (M-step), described next.

\begin{description}
\item[E-step:] Compute the expectation of the complete log-likelihood \eqref{mixMultinomialcomplete}, with
respect to the missing variables $\textbf{Z}$, given the observed data $\textbf{Y}$, and the current parameter estimate $\widehat{\Theta}^{(t)}$ (where $t$ is the iteration counter). Because $\log p(\textbf{Y},\textbf{Z}|\Theta)$ is linear with respect to $\textbf{Z}$ (as is clear in \eqref{mixMultinomialcomplete}),
\begin{equation*}
\mathbb{E}\left[ \log p(\textbf{Y},\textbf{Z}|\Theta) \bigl| \textbf{Y},\widehat{\Theta}^{(t)}\right] = \log p(\textbf{Y},\bar{\textbf{Z}}^{(t)}\bigl| \Theta)
 \end{equation*}
where each element $\bar{z}_{ik}^{(t)}$ of $\bar{\textbf{Z}}^{(t)}$ is given by
\begin{equation}
\bar{z}_{ik}^{(t)} = \mathbb{E}\left[ Z_{ik} \bigl| \textbf{Y},\widehat{\Theta}^{(t)}\right] = \mathbb{P} \left[ Z_{ik} = 1 \bigl| \underline{y}_i,\widehat{\Theta}^{(t)}\right] = \frac{\alpha_k\; f(\underline{y}_i | \underline{\theta}_k)}{\sum_{j=1}^K \alpha_j\; f(\underline{y}_i | \underline{\theta}_j)},\label{Estep_general}
 \end{equation}
since $Z_{ik}$ is binary and conditionally independent from all $\underline{y}_j$, for $j\neq i$; the third equality results simply from Bayes' law.

\item[M-step:]
Update the parameter estimates by maximizing the current estimate of the expected complete log-likelihood
\begin{equation}
\widehat{\Theta}^{(t+1)}=\arg\max_{\Theta}\; \log p(\textbf{Y},\bar{\textbf{Z}}^{(t)}\bigl|\Theta) + \log p(\Theta),
 \end{equation}
where $p(\Theta)$ is a prior, in the case of MAP estimation; in the case of ML estimation, the term $\log p(\Theta)$ is absent.

\end{description}
A more detailed derivation of the EM algorithm, its convergence properties, extensions, and instances for different type of missing-data models, can be found, e.g., in \cite{McLachlan2008}, \cite{Gupta}.

\section{Model selection for categorical data}
Model selection is an important problem in statistical analysis \cite{CeleuxEtAl2014}. In model-based clustering, the term {\it model selection} usually refers to the problem of determining the number of clusters, although it may also refer to the problem of selecting the structure of the clusters.
Model-based clustering provides a statistical framework to solve this problem \cite{FraleyaRaftery2002}, usually resorting to {\it information criteria}. The rationale of such criteria is that fitting a model with a large number of clusters requires estimation of a very large number of parameters and a
consequent loss of precision in these estimates. Therefore, one should penalize excessive model complexity and simultaneously try to increase the model's fit to the data, based on the likelihood function. The best-known information criteria are BIC, AIC, and their modifications, namely  the {\it consistent AIC} (CAIC) \cite{Bozdogan1987}, and the {\it modified AIC} (MAIC) \cite{Bozdogan94}.
Other criteria that have been proposed include the {\it integrated completed likelihood} (ICL) \cite{BiernackiEtAl200}, the {\it minimum description length} (MDL) \cite{Rissanen}, and the {\it minimum message length} (MML) criterion \cite{WB1968}. Information criteria are well-known and easily implemented, the final model being selected according to a compromise between its fit to data and its complexity.

In this work, we use the MML criterion to choose the number of components of a mixture of multinomials. MML is based on the information-theoretic view of estimation and model selection, according to which an adequate model is one that allows a short description of the observations \cite{WB1968}. MML-type criteria evaluate statistical models according to their ability to compress a message containing the data, looking for a balance between choosing a simple model and one that describes the data well.

According to Shannon's information theory, if $Y$ is some random variable with probability distribution $p(y|\Theta)$, the
optimal code-length (in an expected value sense) for an outcome $y$ is
$l(y|\Theta) = - \log_2 p(y|\Theta),$ measured in bits
(from the base-2 logarithm) \cite{CT1991}.
If $\Theta$ is unknown, the total code-length function has two parts:
$l(y,\Theta) = l(y|\Theta) + l(\Theta)$;
the first part encodes the outcome $y$, while the second part encodes
the parameters of the model. The first part corresponds the fit of
the model to the data (better fit corresponds to higher compression), while
the second part represents the complexity of the model.
Different ways to compute $l(\Theta)$ are derived from different statistical frameworks and yield
different flavors of information criteria, namely MDL and MML, where MML admits the existence of a prior $p(\Theta)$, while MDL does not.

The message length function for a mixture of distributions
(as developed in \cite{BaxterOlivier2000}) is:
\begin{equation}
\label{eq:hT}
 l(y,\Theta) = -\log p(\Theta) - \log p(y|\Theta) + \frac{1}{2} \log | I(\Theta) | +                                                                                                       \frac{C}{2} \left(1-\log(12)\right),
\end{equation}
where  $p(y|\Theta)$ is the likelihood function,
$ I(\Theta) \equiv -E \left[ \frac{\partial^2}{\partial{\Theta}^2} \log p(Y|\Theta) \right]$ is the expected Fisher information matrix, $\vert I(\Theta)\vert$ its determinant, and $C$ is the the number of parameters of the model that need to be estimated. For example, for the $K$ mixture multinomial distributions presented in (\ref{mixMultinomial}),
\[
C=(K-1) + K \left( \sum_{l=1}^L(C_l-1)\right) .
\]

The expected Fisher information matrix of a mixture leads to a complex
analytical form of MML which cannot be easily computed.
To overcome this difficulty, Figueiredo and Jain \cite{FJ2002}
replace the expected Fisher information matrix by its complete-data counterpart
$I_c(\Theta) \equiv - E \left[ \frac{\partial^2}{\partial {\theta}^2} \log p(Y,Z|\Theta) \right]$.
Also, they adopt independent Jeffreys' {\it priors} for the mixture parameters.
The resulting message length function is
\begin{equation}
\label{mml function}
l(y, {\Theta}) = \frac{M}{2} \sum_{k:\, \alpha_k>0} \log \left( \frac{n\;  \alpha_k}{12} \right) + \frac{k_{nz}}{2} \log \frac{n}{12}
 + \frac{k_{nz} (M+1)}{2} - \log p(y,{\Theta})
\end{equation}
where $M$ is the number of parameters specifying each component (the dimension of each $\underline{\theta}_k$)
and $k_{nz}$ the number of components with non zero probability (for more details on the derivation of \eqref{mml function}, see \cite{FJ2002}, \cite{BaxterOlivier2000}).

\section{The proposed MML based EM algorithm}
In order to estimate a mixture of multinomials, we propose to use a variant of the EM algorithm (herein termed EM-MML), which integrates both estimation and model selection, by directly minimizing \eqref{mml function}. This algorithm is an extension, to the multinomial case, of the approach developed by in \cite{FJ2002}  for clustering continuous data, based on a Gaussian mixture model.

The algorithm results from observing that \eqref{mml function} contains, in addition to the log-likelihood term, an explicit penalty on the number of components (the two terms proportional to $k_{nz}$), and a term (the first one) that can be seen as a log-prior on the $\alpha_k$ parameters of $\Theta$, that will directly affect the M-step. Finally, notice that in the presence of a set of multinomial observations $\textbf{Y} = \{ \underline{y}_i,  \;    i=1, \dots, n\}$, the log-likelihood $\log p(\textbf{Y}\bigl|\Theta)$ is as given in \eqref{mixMultinomial}.

\begin{description}
\item[E-step:] The E-step of the EM-MML is precisely the same as in the case of ML or MAP estimation, since the generative model for the data is the same.
Since we are dealing with a multinomial mixture, we simply have to plug the corresponding multinomial probability function in \eqref{Estep_general}, yielding

\begin{equation}\label{Estep}
 \bar{z}_{ik}^{(t)} = \frac{\alpha_k\, \prod_{l=1}^L \left[ n_l! \prod_{c=1}^{C_l} \frac{(\widehat{\theta}_{klc}^{(t)})^{y_{ilc}}}{y_{ilc}!} \right]}{\sum_{j=1}^K \alpha_j \, \prod_{l=1}^L \left[ n_l! \prod_{c=1}^{C_l} \frac{(\widehat{\theta}_{jlc}^{(t)})^{y_{ilc}}}{y_{ilc}!} \right]},
\end{equation}
for  $i=1, \dots,n$ and $k=1,\dots,K$.

\item[M-step:] For the M-step, noticing that the
first term in \eqref{mml function} can be seen as the negative log-prior $-\log p(\alpha_k) = \frac{C-K+1}{2K}\log \alpha_k$ (plus a constant), and enforcing the conditions that $\alpha_k \geq 0$, for $k=1,...,K$ and that $\sum_{k=1}^K \alpha_k = 1$, yields the following updates for the estimates of the $\alpha_k$ parameters:
\begin{equation}
\label{Mstep}
\widehat{\alpha}^{(t+1)}_k = \frac{\displaystyle \max \left\lbrace 0, \sum_{i=1}^n \bar{z}_{ik}^{(t)} - \frac{C-K+1}{2K} \right\rbrace }{\displaystyle \sum_{j=1}^K \max \left\lbrace 0, \sum_{i=1}^n \bar{z}_{ij}^{(t)} - \frac{C-K+1}{2K} \right\rbrace },
 \end{equation}
for $k=1,...,K$.
Notice that, some $\widehat{\alpha}^{(t+1)}_k$ may be zero; in that case, the $k$-th component is excluded from the mixture model. The multinomial parameters corresponding to components with $\widehat{\alpha}^{(t+1)}_k = 0$ need not be further calculated, since these components do not contribute to the likelihood.
For the components with non-zero probability, $\widehat{\alpha}^{(t+1)}_k>0$, the estimates of multinomial parameters are updated to their standard weighted ML estimates:
\begin{equation}
\label{est_theta}
\widehat{\theta}^{(t+1)}_{klc} = \frac{\displaystyle \sum_{i=1}^n \bar{z}_{ik}^{(t)} y_{ilc}}{n_l \displaystyle \sum_{i=1}^n \bar{z}_{ik}^{(t)}},
\end{equation}
for  $k=1, \dots,K$, $l=1,\dots,L$, and $c=1,\dots,C_l$. Notice that, in accordance with the meaning of the $\theta_{klc}$ parameters, $\sum_{c=1}^{C_l}\widehat{\theta}^{(t+1)}_{klc} = 1$.

\end{description}

The EM-MML algorithm for clustering categorical data and selecting the number of clusters simultaneously
is summarized in Figure \ref{f:pseudocode}.

\begin{figure*}[h!]
\caption{\small The EM-MML algorithm}
\label{f:pseudocode}
\begin{center}
{\tiny
\begin{tabular}{|ll|}
\hline
\textbf{Input:}
 & data: $\textbf{Y}=\{\underline{y}_i, \; i=1, \dots, n\}$ \\
  &the minimum number of segments: $K_{min}$ \\
  &the maximum number of segments: $K_{max}$ \\
  & minimum increasing threshold for the log-likelihood function:  $\delta$\\
 \textbf{Ouput:}
 & number of segments: $K$ \\
 & segments probabilities: $\left\{ \alpha_1,\dots,\alpha_K\right\}$ \\
 & multinomial parameters:$\{\underline{\theta}_1, \dots, \underline{\theta}_K\}$\\
 & \\
  \textbf{Initialization:} 
 &initialization of the parameters resorts to the empirical distribution:\\
 & \hspace{0.5cm} $p(\underline{y}_l \vert \underline{\theta}_{lk})$ , $(l=1, \dots,L$ ; $k=1, \dots,K_{max})$\\
 & set the segment's probability:
  $\alpha_k = 1/K_{max}$  $(k=1, \dots,K_{max})$ \\
 & store the initial log-likelihood \\
 & store the initial message length $(iml)$ \\
 & $minml$ $\leftarrow iml$ \\
 \multicolumn{2}{|l|}{continue $\leftarrow 1$} \\
 \multicolumn{2}{|l|}{\cwhile continue $= 1$ \cdo} \\
 \multicolumn{2}{|l|}{\hspace{0.5cm} \cwhile increases on log-likelihood are above $\delta$ \cdo} \\
 \multicolumn{2}{|l|}{\hspace{1cm} $k=K_{max}$ }\\
 \multicolumn{2}{|l|}{\hspace{1cm} \cwhile $k \geqslant K_{min}$ \cdo} \\
 \multicolumn{2}{|l|}{\hspace{1.5cm} compute $\widehat{\alpha}_k$ according to (\ref{Mstep})}\\
 \multicolumn{2}{|l|}{\hspace{1.5cm} \cif $\widehat{\alpha}_k = 0$} \\
 \multicolumn{2}{|l|}{\hspace{2cm} remove the component $k$ of the mixture} \\
 \multicolumn{2}{|l|}{\hspace{2cm} $K_{max}$ $\leftarrow K_{max}-1$}\\
 \multicolumn{2}{|l|}{\hspace{2cm}$\left\{ \widehat{\alpha}_1,\dots,\widehat{\alpha}_{K_{max}}\right\}$  $\leftarrow$ $\left\{ \frac{ \widehat{\alpha}_1}{\sum_{k=1}^{k_{max}} \widehat{\alpha}_k} ,\dots,\frac{\widehat{\alpha}_{K_{max}}}{\sum_{k=1}^{k_{max}} \widehat{\alpha}_k}\right\}$ }\\
 \multicolumn{2}{|l|}{\hspace{1.5cm}\celse}\\
 \multicolumn{2}{|l|}{\hspace{2cm} compute $\widehat{\theta}$ according to (\ref{est_theta})} \\
 \multicolumn{2}{|l|}{\hspace{2cm} E-step according to (\ref{Estep})} \\
 \multicolumn{2}{|l|}{\hspace{2cm} $k$ $\leftarrow k-1$ }\\
 \multicolumn{2}{|l|}{\hspace{1.5cm}\cendif} \\
 \multicolumn{2}{|l|}{\hspace{1cm} \cendwhile} \\
 \multicolumn{2}{|l|}{\hspace{1cm} compute log-likelihood} \\
 \multicolumn{2}{|l|}{\hspace{1cm} compute message length (ml)} \\
 \multicolumn{2}{|l|}{\hspace{0.5cm} \cendwhile} \\
 \multicolumn{2}{|l|}{\hspace{0.5cm} \cif $ml<minml$} \\
 \multicolumn{2}{|l|}{\hspace{1cm} $minml \leftarrow ml$ } \\
 \multicolumn{2}{|l|}{\hspace{1cm} segment's probability  $\leftarrow$ current parameters} \\
 \multicolumn{2}{|l|}{\hspace{1cm} multinomial probabilities  $\leftarrow$ current parameters} \\
 \multicolumn{2}{|l|}{\hspace{0.5cm}\cendif} \\
 \multicolumn{2}{|l|}{\hspace{0.5cm} \cif $K_{max}>K_{min}$} \\
 \multicolumn{2}{|l|}{\hspace{1cm} $k_{remove}=arg min \left\{ \alpha_1,\dots,\alpha_K\right\}$ } \\
 \multicolumn{2}{|l|}{\hspace{1cm} remove the component $k_{remove}$ of the mixture model} \\
 \multicolumn{2}{|l|}{\hspace{1cm} $K_{max}$ $\leftarrow K_{max}-1$} \\
 \multicolumn{2}{|l|}{\hspace{1cm} $\left\{ \alpha_1,\dots,\alpha_{K_{max}}\right\}$  $\leftarrow$ $\left\{ \frac{ \widehat{\alpha}_1}{\sum_{k=1}^{k_{max}} \widehat{\alpha}_k} ,\dots,\frac{\widehat{\alpha}_{K_{max}}}{\sum_{k=1}^{k_{max}} \widehat{\alpha}_k}\right\}$ }\\
 \multicolumn{2}{|l|}{\hspace{0.5cm}\celse} \\
 \multicolumn{2}{|l|}{\hspace{1cm} continue $\leftarrow 0$} \\
 \multicolumn{2}{|l|}{\hspace{0.5cm} \cendif} \\
 \multicolumn{2}{|l|}{ \cendwhile} \\
\multicolumn{2}{|l|}{The best solution corresponds to the minimum message length obtained. } \\
\hline
\end{tabular}
}
\end{center}
\end{figure*}

\section{Data analysis and results}
\subsection{Evaluating the EM-MML performance on synthetic data}
To evaluate the performance of the EM-MML algorithm, we begin by resorting to synthetic data sets: 2-components mixtures of multinomials (10 data sets) and 3-components mixtures (10 data sets) are used.
In order to provide useful insights regarding the EM-MML performance, the 10 data sets within each setting exhibit diverse degrees of cluster separation. For this purpose, we measure the separation between all pairs of clusters $(C^k;C^{k'})$ forming in a partition $\Pi^K$ according to
\begin{equation}\label{separation}
\mbox{separation}(\Pi^K)= \frac{2}{k(k-1)} \sum_{k \neq k'} \frac{1}{2} \left[ D_{KL}(C^k;C^{k'}) + D_{KL}(C^{k'};C^k) \right],
\end{equation}
where
\[
D_{KL}(C^{k};C^{k'})=\sum_{l=1}^{L} \sum_{c=1}^{C_l} P(Y_{l}^{k}=c) \log \frac{P(Y_{l}^{k}=c)}{P(Y_{l}^{k'}=c)} = \sum_{l=1}^{L} \sum_{c=1}^{C_l}
\widehat{\theta}_{klc}\log \frac{\widehat{\theta}_{klc}}{\widehat{\theta}_{k'lc}}
\]

is the sum of the Kullback-Leibler divergence between the $L$ multinomial distributions corresponding to components $k$ and $k'$ of the mixture. This symmetric measure of separation ranges from around
$0.01$ (poor clusters' separation) to $0.17$ (good clusters' separation), in the experimental scenarios.

\begin{figure}[H]
  \includegraphics[width=0.5\textwidth]{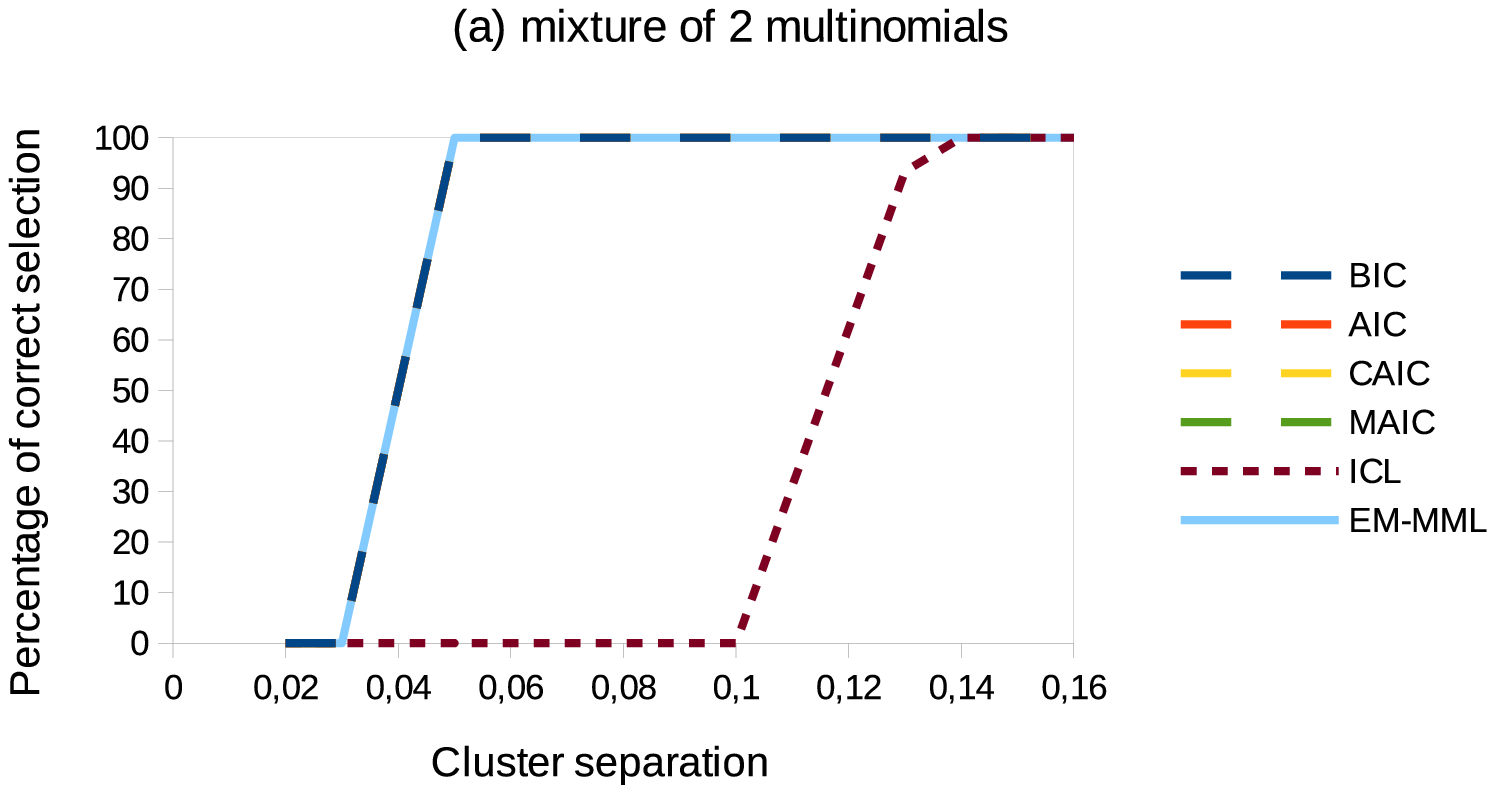}
  \includegraphics[width=0.5\textwidth]{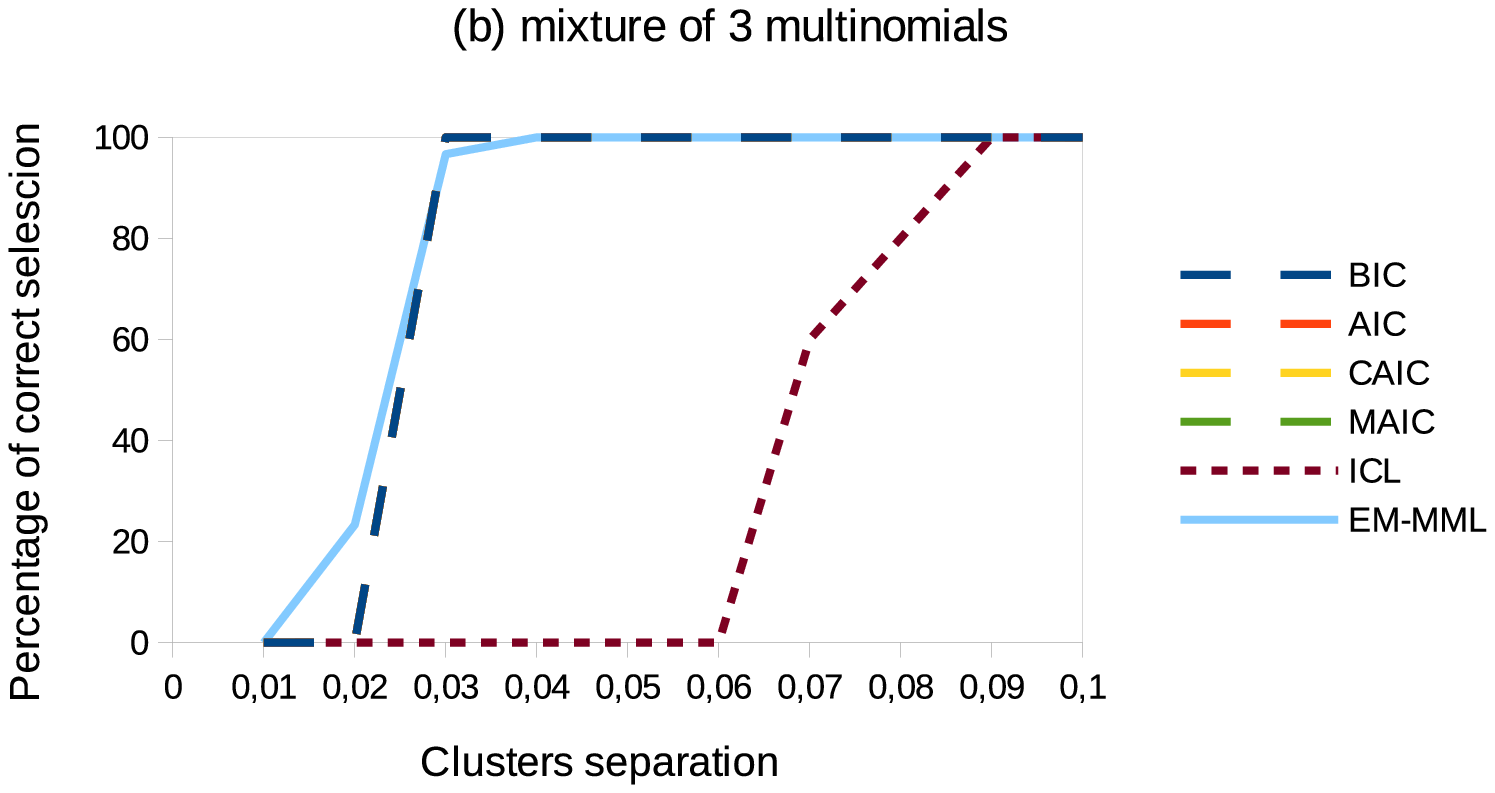}
  \caption{Comparison of several model selection criteria on synthetic data
  from mixtures of multinomials with $K=2$ (left) and $K=3$ (right).}
  \label{fig:cluster-separation}
\end{figure}

The EM-MML results are compared with those obtained from a standard EM algorithm and well-known information criteria, namely  BIC, AIC, CAIC, MAIC, and ICL. The performance of the various criteria and methods is assessed by the rate (over 30 runs) of correct selection of the true number of clusters, and also by the computation time. The results are presented in Figure~\ref{fig:cluster-separation}.

In the two-components data sets  with separation lower than 0.04,
all the approaches identify only one cluster. ICL is unable to recover the true number of clusters, even with moderate separation (up to around 0.12). The other criteria, and our EM-MML approach, correctly identify two clusters
when separation is above 0.04. For three-components mixtures, the ICL criterion is able to identify the correct number of clusters only for cluster separation higher than 0.08. The other criteria (including EM-MML) present similar results identifying three clusters for separations larger than 0.03. Generally one can thus conclude that the EM-MML has a similar performance to BIC, AIC, CAIC, and MAIC in recovering the true number of clusters, while ICL clearly underperforms the other methods for not clearly separated components.

In order to further evaluate the comparative performance of EM-MML, we compare its computation times with that of BIC. Notice that BIC and other information criteria have similar computation times. Based on a 300 runs sample from 2-components mixtures of multinomials, we obtain the following average computation times: 146.84 seconds for EM-MML and 230.67 seconds for BIC. A paired-samples t test yields t = -17.06; df = 299 and p-value $< 0.01$, allowing  to conclude that EM-MML is significantly faster than BIC. For the  3-components mixtures of multinomials, the average of computation times are 194.38 and 239.27 seconds for EM-MML and BIC respectively. Again, the null hypothesis of a paired-samples t test is rejected (t = -6.88; df = 299 and p-value $< 0.01$), showing that EM-MML is significantly faster than BIC. Overall, the EM-MML method shows good performance when dealing with synthetic data sets: when selecting the true number of clusters it has similar performance to BIC, AIC, CAIC, and MAIC, and outperforms ICL. In terms of computation time, since EM-MML does not require a sequential approach,
it becomes clearly faster than the other criteria.

\subsection{Experiments on real data}
Additional insight into the performance of EM-MML is obtained by applying it to two real data sets from the European Social Survey (ESS). ESS is a biennial survey started in 2002, which measures the attitudes, beliefs, values, and behaviour patterns of European populations. The most recent survey was in 2012 (round 6), covering 30 countries and 243 regions.

For the purpose of our experiment, we aggregate the ESS data by region,
taking into account sampling weights (ESS weights) which are meant to provide the sample representativeness.
Clustering is performed based on: 1) the {\it trust in} some institutions (namely, in the
country's parliament, legal system, police, politicians, political parties, the
European Parliament, the United Nations); 2) the {\it satisfaction with} life as whole, the economy, the government, and the functioning of the democracy.
We recode responses into binary variables: distrust/trust and dissatisfied/satisfied.

The summary of the comparisons of the several model selection criteria on the two ESS data sets is presented in Tables~\ref{Trust VCramer} and~\ref{How Satisfied VCramer}. To measure the relationship between uncovered segments and the clustering base variables, we resort to the Cramer's V association measure
- which ranges from 0 (no association) to 1 (perfect association) - and to its sum as a proxy of
variables' discriminant capacity.

\begin{table}[h!]
\caption{Comparison of Cramer's V association between segmentation base variables ``trust in" and the segments obtained with each criterion.}
\label{Trust VCramer}
\small{
\begin{tabular}{lcccccc}
  \hline
      &  BIC & AIC & CAIC & MAIC & ICL & EM-MML \\
      Number of segments & 11& 17& 11& 18& 11& 4\\
      Trust in & & & & & &\\
  \hline
    country's parliament & .58	& .53 &.58 &.55 &.58 &.72\\
    the legal system & .55	& .54	& .55	& .57	& .55	& .67 \\
	the police & .52	& .51	& .52	& .50	& .52	& .57 \\
	politicians & .64	& .57	& .64	& .58	& .64	& .78 \\
	political parties & .64	& .60	& .64	& .58	& .64	& .78 \\
	the European Parliament & .46	& .45	& .46	& .47	& .46	& .52 \\
	the United Nations& .47	& .43	& .47	& .48	& .47	& .52 \\
 \hline
 Sum & 3.86	& 3.63	& 3.86	& 3.73	& 3.86& 	4.56 \\
   \hline
\end{tabular} }
\end{table}

\begin{table}[h!]
\caption{Comparison of sum of Cramer's V association between segmentation base variables "satisfaction with" and the segments obtained with  each criterion.}
\label{How Satisfied VCramer}
\small{
\begin{tabular}{lcccccc}
  \hline
      &  BIC & AIC & CAIC & MAIC & ICL & EM-MML \\
      Number of segments & 13& 18& 13& 18& 13& 7\\
      How satisfied with & & & & & & \\
      \hline
      life as a whole & .59 & .55 & .59 & .55 & .59 & .55  \\
      present state of economy in country & .58 & .55 & .58 & .55 & .58  & .62 \\
      the national government & .56 & .55 & .56 & .55 & .56  & .60\\
      the way democracy works in country & .56 & .53 & .56 & .53 &  .56 & .63\\
       \hline
      Sum & 2.29 & 2.18 & 2.29 & 2.18 & 2.29 & 2.40 \\
   \hline

\end{tabular}}
\end{table}

The number of segments selected by the EM-MML is much lower than for the remaining
criteria. This fact avoids estimation problems associated with very small segments and also improves the interpretability of the clustering solution. In addition,
the total value of the Cramer’s V association measure (which evaluates the relationship between clusters and clustering base variables) is higher for EM-MML,
indicating variables with higher discriminant capacity in the EM-MML solution. The segments selected by EM-MML criterion are presented in Tables \ref{Trust MML} and \ref{How Satisfied MML}.

\begin{table}[h!]
\caption{EM-MML segmentation of the "trust in" data. }
\label{Trust MML}
\small{
\begin{tabular}{lcccc}
  \hline
    Trust in  &  Segment 1 & Segment 2 & Segment 3 & Segment 4 \\
  \hline
	country's parliament & 59.2\%  &  18.2\%  & 14.2 \%  &  35.8\% \\
	the legal system & 72.0\%  &  23.2\%  &  17.7\%  &  55.4\% \\
	the police & 81.4\%   &  41.6\%  &  20.8\%  &  72.9\% \\
	politicians &  45.6\%   &  10.6\%  &  8.7\%  &  20.8\% \\
	political parties & 44.1\%  &  10.8\%  &  9.9\%  &  19.1\% \\
	the European Parliament & 41.7\%  &  26.9\%  &  20.2\%  &  28.7\% \\
	the United Nations & 64.7\%   &  35.0\%  &  24.6\%  &  45.0\% \\
   \hline

\end{tabular}
}
\end{table}

\begin{table}[h!]
\caption{EM-MML segmentation of the "satisfaction with" data.}
\label{How Satisfied MML}
\small{
\begin{tabular}{lccccccc}
  \hline
     How satisfied with &  Seg. 1 & Seg. 2 & Seg. 3 & Seg. 4 & Seg. 5 & Seg. 6 & Seg. 7\\
  \hline
	 life as a whole & 92.5\% & 93.0\% & 29.3\% & 79.7\% & 71.8\% & 59.6\% & 88.0\% \\
	 present state of economy & 85.8\% & 36.4\% & 3.5\% & 20.7\% & 35.5\% & 15.8\% & 57.3\% \\
	 in country\\
	the national government & 70.5\% & 56.7\% & 8.5\% & 23.4\% & 34.5\% & 21.9\% & 42.9\% \\
	the way  democracy& 85.8\% & 81.3\% & 9.8\% & 47.1\% & 46.1\% & 22.8\% & 68.2\% \\
	works in country\\
  \hline

\end{tabular}
}
\end{table}

The "trust in institutions" EM-MML segmentation yields four segments. For example, Lisbon is in segment 2, where all the regions have low trust in institutions. The lowest trust values, in this segment, refer to trust in politicians and political parties. Comparing the trust in different institutions, Lisbon citizens have moderate trust in the country police. Dublin, another European capital city, is in segment 4, characterized by a  higher trust in institutions, specially in the police (72.9\%) and
the legal system (55.4\%). The {\it separation} of these segments (according to the measure in \eqref{separation}) is 1.46, indicating that they are well separated.

The seven EM-MML segments for "satisfaction with" are also well separated segments (separation$(\Pi^7) =1.26$). In these segmentation, Lisbon is in segment 6 and Dublin in segment 4. In both capitals, people only feel satisfied with life as a whole, but in Dublin they feel generally more satisfied than in Lisbon.

\section{Discussion and perspectives}
In this work, a model selection criterion and method for finite mixture models of categorical observations was proposed. The new algorithm simultaneously performs  model estimation and selects the number of components/clusters.
When compared to information criteria, which are commonly associated with the use of the EM algorithm, the EM-MML method exhibits several advantages: 1) it easily recovers the true number of clusters in synthetic data sets with various degrees
of separation; 2) its computations times are significantly lower than those required by the sequential use of EM in standard approaches (such as BIC); 3) when applied to real data sets it produces more parsimonious solutions, thus easier to interpret.
An additional advantage of the proposed approach that stems from obtaining more parsimonious solutions is that such solutions have a higher number of observations per cluster, thus helping to overcome eventual estimation problems.

Since the performance of the EM-MML is encouraging for selecting the number of clusters, and the same criterion was already used for feature selection \cite{LFJ2004}, future developments will include the integration of both model and feature selection.

\bibliographystyle{gSCS}
\bibliography{biblio}

\end{document}